\documentclass[12pt]{article}
\usepackage[utf8]{inputenc}
\usepackage[svgnames]{xcolor}
\usepackage[a4paper, inner=1cm, outer=1cm, top=1cm, bottom=2cm]{geometry}
\usepackage[bottom, multiple]{footmisc}
\usepackage{graphicx}
\usepackage{tikz}
\usetikzlibrary{decorations.markings}
\usetikzlibrary{decorations.pathmorphing}
\usepackage{framed}
\usepackage{csquotes}
\definecolor{shadecolor}{rgb}{0.90,0.90,0.90}
\usepackage{cite}
\usepackage{hyperref}
\usepackage{subcaption}
\usepackage{pifont}
\usepackage{setspace}
\usepackage{amsmath, amssymb, amsthm, float, graphicx,amsfonts,braket}
\numberwithin{equation}{section}
\usepackage{amsthm}

\theoremstyle{definition}

\hypersetup{colorlinks=true, linkcolor=Maroon, citecolor=FireBrick,urlcolor=Green,linktocpage}

\def\beq{\begin{eqnarray}}\def\eeq{\end{eqnarray}}
\def\be{\begin{equation}}\def\ee{\end{equation}}

\def\r{\rho}

\def\m{\mu}

\def\a{\alpha}

\def\b{\beta}

\def\t{\tau}
\def\D{\Delta}
\def\Dphi{\Delta_\phi}

\def\w{\omega}

\def\mW{{\mathcal{W}}}

\usepackage{bm}
\usepackage{bm}

\begin{document}

\title{\bf A Celestial Route to AdS Bulk Locality}
\date{}
\author{Faizan Bhat$^{\a}$\footnote{faizanbhat@iisc.ac.in} ~and Ahmadullah Zahed$^{\a}$\footnote{ahmadullah@iisc.ac.in
}\\~~~~\\
\it ${^\a}$Centre for High Energy Physics,
\it Indian Institute of Science,\\ \it C.V. Raman Avenue, Bangalore 560012, India.}
\maketitle
\vskip 2cm
\abstract{We prove a precise form of AdS bulk locality by deriving analytical two-sided bounds on bulk Wilson coefficients. Our bounds are on the Wilson coefficients themselves, rather than their ratios, as is typically found in the literature. Inspired by the Celestial amplitudes program in flat space, we perform a Celestial transform of the CFT Mellin amplitude of four identical scalars. Using the crossing symmetric dispersion relation (CSDR), we express the resulting amplitude in terms of crossing symmetric conformal partial waves. The partial waves satisy remarkable positivity properties which along with the unitarity of the CFT prove sufficient to derive the bounds. We then employ our methods in the limit of large AdS radius and recover known bounds on flat space Wilson coefficients and new bounds on their large radius corrections. We check that the planar Mellin amplitude of four stress-tensor multiplets in $\mathcal{N} = 4$ SYM satisfies our bounds. Finally, using null constraints, we derive a form of low-spin dominance in AdS EFTs. We find that the low-spin dominance is strongest in flat space and weakens as we move away from flat space towards higher AdS curvature.}

\tableofcontents

\onehalfspacing

\section{Introduction}
What is the space of low-energy effective field theories (EFTs) that can give rise to a consistent theory of quantum gravity in the UV?  Over the last few years, significant progress has been made towards answering this question by deriving sharp two-sided bounds on (usually the ratios of) the Wilson coefficients that parametrize the low-energy EFTs. This has been made possible through the use of twice-subtracted dispersion relations that follow from the assumptions of analyticity and Regge boundedness of the scattering amplitudes\cite{SMSBoot, mandelstam,nuss,Martin, Sboot, ABAS, andrea , dual, piotr,Meltzer,nimayutin2, AnindaKnot, Mizera, Paulos:2020zxx, anant,PHAS,Haring:2022cyf}. Dispersion relations express the low-energy Wilson coefficients in terms of the UV data of the theory. Demanding that scattering amplitudes obey crossing symmetry and unitarity in the UV leads to the the bounds on Wilson coefficients \cite{nima,tolley,SCHFlat,RMTZ,rattazzi,Wang:2020jxr,nimayutin,yutin2,sasha,Davis:2021oce, Caron-Huot:2022ugt, EliasMiro:2022xaa, deRham:2021bll, vichi, Chowdhury:2021ynh}. While the procedure is straightforward without gravity, inclusion of the graviton presents an obstacle due to a pole that arises in the forward scattering limit of the graviton exchange term. This was recently overcome in both flat space and AdS space by measuring the Wilson coefficients in the small impact parameter limit ( see \cite{Swamp, chuotAdS}, also see \cite{tolley2, Kundu:2021qpi}). \\
\\
In this paper, we consider the 2-2 scattering of identical scalars of mass $m \sim 1/R$ in a weakly-coupled gravitational EFT in AdS ($R$ is the AdS radius). We take the view that the full theory of quantum gravity in AdS is defined non-perturbatively by its CFT dual. AdS scattering amplitudes are then naturally expressed in terms of the Mellin representation of the dual CFT correlator \cite{Fitzpatrick:2011ia}. For our case, we therefore consider the Mellin correlator of identical scalar primaries of dimension $\Delta_{\phi} \sim m R$. We assume the following expansion of the Mellin amplitude
\be\label{eq:Wpqdef}
M(s_1,s_2) \sim 8 \pi G M_{\text{gravity}}+ \mathcal{W}_{0,0} + \mathcal{W}_{1,0} x +\mathcal{W}_{0,1}y +\mathcal{W}_{2,0} x^2 +\mathcal{W}_{1,1} x y+ \dots \,
\ee
where $x = -(s_1 s_2 + s_2 s_3+s_3 s_1)$, $y = -s_1 s_2 s_3$ and $s_i$ are related to the usual Mellin-Mandelstam variables by a shift such that $s_1 + s_2 + s_3 =0$. The coefficients $\mathcal{W}_{p,q}$ provide a parametrization of the low-energy EFT in AdS and we shall refer to them as the AdS Wilson Coefficients in this paper. $8 \pi G M_{\text{gravity}}$ is the term corresponding to the graviton exchange in AdS and leads to the familar graviton pole term $\sim 8 \pi G \dfrac{x^2}{y}$ in the flat-space limit. Our goal is to derive two-sided bounds on $\mathcal{W}_{p,q}$. This is in contrast to previous studies\cite{Swamp, chuotAdS} where bounds are typically derived on the ratios of Wilson coefficients or ratio with respect to $G$.
\\ \\
An essential tool in our work is the upliftment of the Celestial CFT techniques for flat space employed in \cite{SGPRAS} to AdS. For details of the flat space Celestial amplitudes program, see \cite{Pasterski:2021raf,McLoughlin:2022ljp,Pasterski:2016qvg,Pasterski:2017kqt,Strominger:2017zoo,Pasterski:2021rjz,Raclariu:2021zjz,Fan:2019emx,Banerjee:2020kaa,Banerjee:2020zlg,Banerjee:2020vnt, MizeraPasterski}. We begin by writing down a twice-subtracted crossing symmetric dispersion relation for the Mellin correlator \cite{RGASAZ} in terms of the celestial variable $z$ and $\w$ defined as
\be 
s_1 = \w^2, \quad s_2 = \-z w^2, \quad s_3 = (z-1)\w^2
\ee
The price one has to pay for having a fully crossing symmetric expansion is the loss of manifest locality. The crossing symmetric conformal partial waves have spurious poles. These poles must cancel when summing over the spectrum and spins. Otherwise, there will be terms with poles in $x$ in \eqref{eq:Wpqdef} which cannot come from a local bulk Lagrangian. Therefore, we must impose the ``locality condition", $\mathcal{W}_{-r,q} =0, \forall~ r > 0$, where $\mathcal{W}_{-r,q}$ is the coefficient of the term $x^{-r}y^q$. We can separate out the spurious pole contribution from each conformal partial wave; the new partial waves are fully crossing symmetric and local, and are called Witten blocks in the literature \cite{Pol,RGAS, PFKGASAZ}.\\ 
\\
We perform a Celestial transform of the CFT Mellin amplitude of four identical scalars (see \eqref{MellinDef}) as follows
\be
\label{CelAmpDef}
 \widetilde{M}(\b,z) = \int_0^{\infty} d\w~ \w^{\b-1} M( \w^2, -z \w^2)
\ee
The amplitude $\widetilde{M}(\b,z)$ is known to have poles at $\beta = -2n$ i.e. negative even integers. The residues at these poles capture all the information about the low-energy Wilson Coefficients \cite{NimaCel} . More precisely, we have
\be
\label{eq:LowEnergyRes}
\text{Res}_{\b = -2n} \widetilde{M}(\b,z) \equiv \widetilde{\mathcal{W}}(n,\r) = \sum_{\substack{p,q \\ 2p+3q=n}}\frac{(-1)^q}{2^{p+q}}\mathcal{W}_{p,q}(1-\rho)^p (1+\rho)^q
\ee 
where we use the variable $\rho (z) = -1 +2z -2z^2$ introduced in \cite{SGPRAS}. Using the dispersion relation, we write an explicit expression for the residue $\widetilde{\mathcal{W}}(n,\r)$ in terms of a sum over Celestial Witten blocks (see \eqref{mtildebetaz1}) for each $n$. {We examine the properties of these blocks in the variable $\rho$ and find that they satisfy remarkable positivity properties, namely sign definiteness and typically realness. The Celestial transform plays an important role here as these properties of the Celestial Witten blocks in the $\rho$ variable are not obvious without the Celestial transform. These properties had not been demonstrated for Mellin amplitudes before, mainly due to the appearance of Mack polynomials in Mellin space partial wave expansions which are more complicated than the Gegenbauer polynomials that appear in the flat-space partial wave expansions.} The properties are sufficient for us to derive two-sided bounds on all Wilson coefficients except when $n =2, 3$, i.e. for $\mW_{1,0}$ and $\mW_{0,1}$. For these Wilson coefficients, the results of \cite{Swamp, chuotAdS} suggest that only the ratio $\mW_{0,1}/ \mW_{1,0}$ is expected to have two-sided bounds. In an upcoming work \cite{FBAZ_GFT}, we will derive precise two-sided  bounds on $\mW_{0,1}/ \mW_{1,0}$. These will follow from positivity properties of the full Mellin amplitude which imply two-sided bounds on ratios of all Wilson coefficients in a CFT.\\ 
\\
The positivity properties we find for Celestial Witten blocks display an interesting pattern. For $n \ge 4$, we find that they hold beyond a critical spin $\ell_c$ only above some critical twist $\t^*(n,\ell)$. For example, for $n=4$ (i.e. for $\mW_{2,0}$), we find that positivity holds for $\ell \ge 4$ only above some twist $\t^*(4,4)$ which we can compute numerically. For higher $n$, there is a similar critical twist for spins greater than four and in general we find that $\t^*(n,\ell)$ grows with $n$. In all, this means that to bound the Wilson coefficients via our procedure, it is sufficient that all spins $\ge 4$ exist above a certain twist $\text{max}(\t^*(n,\ell))$. { The bounds we derive are given in terms of discrete moments of OPE coefficients (see (4.4)) of operators with spin $\le 2$. This demonstrates in a precise fashion, a form of the HPPS conjecture made in \cite{HPPS} where similar conditions were conjectured to be sufficient for a (large N) CFT to have a local AdS dual. Our proof of the HPPS conjecture aligns with the findings of \cite{chuotAdS} where two-sided bounds were derived on the ratios of Wilson coefficients.}\\
\\
Next, to make contact with known examples, we consider the Witten block expansion in the flat-space limit. We show that in the flat-space limit \cite{Penedones:2010ue}, our bounds reproduce the two-sided bounds on the flat-space Wilson coefficients \cite{SGPRAS}. In some cases, we are also able to bound the sub-leading $1/R^2$ corrections to the flat-space Wilson coefficients. As a check of our results, we  demonstrate that the AdS Virasoro-Shapiro amplitude (see \cite{Alday1, Alday2}) expanded in the large $R$ limit satisifies our bounds up to order $1/R^2$.\\
\\
We end with a discussion of low-spin dominance in AdS EFTs. This refers to the case when the contribution of low-spin (spin 0 and spin 2) partial waves to the scattering amplitude dominates over higher spins. Through the use of null constraints, we are able to quantify low-spin dominance. Our results suggest that it is the strongest in the flat space limit and weakens as we move towards higher AdS curvature (see figure \eqref{fig:LSDPLOT}).
\\ \\
The paper is organized as follows. In section 2, we review the crossing symmetric dispersion relation (CSDR) for Mellin amplitudes. In section 3, using the CSDR, we write down a partial wave expansion for the Celestial-transformed Mellin amplitude in terms of the Celestial Witten blocks. We then study the positivity properties of the Celestial Witten blocks in section 4 and show that they naturally lead us to focus on CFTs satisfying the HPPS conjecture. We use the positivity properties to derive two-sided bounds on AdS Wilson coefficients. In section 5, we work in a large $R$ expansion and find known bounds on flat space Wilson coefficients and new bounds on their sub-leading $1/R^2$ corrections. We check that the AdS Virasoro-Shapiro amplitude satisfies our bounds. Section 6 discusses low-spin dominance in AdS EFTs and in section 7, we conclude with a discussion of our results and future work.

\section{Crossing symmetric Representation of Mellin Amplitudes: A brief review}\label{crossreview}
The Mellin amplitude $M\left(s_{1}, s_{2}\right)$ (following the conventions of \cite{RGASAZ}) is defined via
\be
\label{MellinDef}
\mathcal{G}(u, v)=\int_{c- i\infty}^{c+ i\infty} \frac{d s_{1}}{2 \pi} \frac{d s_{2}}{2 \pi} u^{s_{1}+\frac{2 \Delta_{\phi}}{3}} v^{s_{2}-\frac{\Delta_{\phi}}{3}} \mu\left(s_{1}, s_{2}, s_{3}\right) M\left(s_{1}, s_{2}\right)\,
\ee
where $\mathcal{G}(u,v)$ is the position space conformal correlator of four identical scalars of dimension $\Delta_{\phi}$. The measure factor $\m(s_1,s_2,s_3)=\Gamma^{2}\left(\frac{\Delta_{\phi}}{3}-s_{1}\right) \Gamma^{2}\left(\frac{\Delta_{\phi}}{3}-s_{2}\right) \Gamma^{2}\left(\frac{\Delta_{\phi}}{3}-s_{3}\right)$. The amplitude satisfies full crossing symmetry, i.e, $M\left(s_{1}, s_{2}\right) = M\left(s_{2}, s_{3}\right)=M\left(s_{3}, s_{1}\right)$. In \cite{RGASAZ}, using a crossing symmetric dispersion relation, it was shown that the amplitude can be expressed through a fully crossing symmetric conformal partial wave expansion as follows
\be
\label{CSExpAmp}
\begin{split}
M(s_1,s_2)&=\alpha_{0}+\sum_{\t, \ell, k}^{\infty} \frac{c_{\t, \ell}^{(k)}}{\tau_{k}}  \mathcal{P}_{\t,\ell}^{(k)}\left(\sqrt{\frac{\tau_k+3a}{\tau_k-a}}\right)\left(\frac{s_1}{\tau_{k}-s_1}+\frac{s_2}{\tau_{k}-s_2}+\frac{s_3}{\tau_{k}-s_3}\right), \quad a = \frac{y}{x}
\end{split}
\ee
where $x = -(s_1 s_2 + s_2 s_3+s_3 s_1)$ and $y = -s_1 s_2 s_3$. The sum runs over the twist $\t = \D -\ell$ and spin $\ell$ of the primaries, and the $k$ sum gives the contribution from descendants. $\a_0$ is a subtraction constant and $\tau_k=\frac{\tau}{2}+k-\frac{2\Dphi}{3}$.  $c_{\t, \ell}^{(k)}$ is proportional to the OPE coefficient squared (see Appendix \eqref{AppA} for details). We consider $a$ real and  $-\frac{\tau^{(0)}}{3}\leq a<\frac{2\tau^{(0)}}{3}$( see \cite{RGASAZ}), where $\tau^{(0)} = min\left(\tau_k \right)$.  We further define
\be
\mathcal{P}_{\t,\ell}^{(k)}\left(x=1+\frac{2s_2^{'}(\tau_k,a)}{\tau_k}\right)=\mathcal{P}_{\t,\ell}^{(k)}\left(\sqrt{\frac{\tau_k+3a}{\tau_k-a}}\right)\equiv P_{\t+ \ell, \ell}\left(\tau_{k}, s_{2}^{\prime}\left(\tau_{k}, a\right)\right)\,,
\ee
where $P_{\Delta, \ell}(s_1,s_2)$ is the Mack Polynomial (see Appendix \eqref{AppA}) and $s'_{2}\left(\tau_{k}, a\right)=-\frac{\tau_{k}}{2}\left[1-\left(\frac{\tau_{k}+3 a}{\tau_{k}-a}\right)^{1 / 2}\right]$. For a more detailed discussion of CSDR, we refer the reader to \cite{AK, ASAZ, AZreview, DCPHAZ, Bissi:2022fmj} and for properties of Mellin amplitudes--see  \cite{joaopaper, joaopaper2, Bissi:2022mrs}. 

\section{The Celestial Witten Block expansion}
The amplitude $\widetilde{M}(\beta, z)$ defined in \eqref{CelAmpDef} is known to have poles in $\b$ at negative even integers i.e $\beta =-2n$ \cite{NimaCel}. The residue at these poles is related to the low-energy expansion of the amplitude as given in \eqref{eq:LowEnergyRes}. Using \eqref{CSExpAmp}, we can express the residue in a partial wave expansion as follows
\be
\label{mtildebetaz1}
\begin{split}
& \mathrm{Res}_{\beta= -2 n} \widetilde{M}(\beta, z) = (-1)^n  \sum_{\t,\ell,k} { \tau_k^{-n-1}{c}^{(k)}_{\t, \ell}}\bigg[ (-1)^n  \mathcal{P}_{\t,\ell}^{(k)}\left( 1-2 z \right) + z^{n} \mathcal{P}_{\t,\ell}^{(k)}\left(  \frac{z-2}{z}  \right)  \\
& +   (1-z)^{n} \mathcal{P}_{\t,\ell}^{(k)}\left(  \frac{z+1}{z-1}  \right) +   (z(1-z))^{n-\ell} (z^{2}-z+1)^{-n+3}  \mathcal{Q}_{\t,\ell}^{(k)}(-2n,z) \bigg]
\end{split}
\ee
The $\mathcal{Q}_{\t,\ell}^{(k)}(\beta,z)$ are polynomials in $z$ given as
\be 
\begin{split}
  \mathcal{Q}_{\t,\ell}^{(k)}(\beta,z) &= (z(1-z))^\ell (z^2-z+1)^{-3} \sum\limits_{i=1}^3 \mathcal{Z}_{\t,\ell}^{(k)}(\beta,x_{i})\,, \\
 \mathcal{Z}_{\t,\ell}^{(k)}(\beta, x_{i}) &= \frac{e^{- i \pi \beta/2}}{\left(\frac{\ell}{2}-1\right)!} \  \frac{d^{\ell/2-1}}{dx^{\ell/2-1}}\bigg[ \frac{x^{\beta/2} (1+x)^{\ell/2}}{(x-x_{i})} \mathcal{P}_{\t,\ell}^{(k)}\left( \sqrt{\frac{1-3x}{1+x}}\right) \bigg] \bigg|_{x=-1} 
\end{split}
\ee
with $x_{1} = -z(z-1)(z^{2}-z+1)^{-1}, \ x_{2} = (z-1)(z^{2}-z+1)^{-1}, \ x_{3} = -z(z^{2}-z+1)^{-1}$.

\subsubsection*{Null/Locality Constraints}
As discussed in the Introduction, the crossing symmetric partial waves have spurious poles which must cancel while summing over the spectrum and spins \cite{RGASAZ, SGPRAS}\footnote{In the fixed $t$ dispersion relation, there are no such negative powers and the corresponding sum rules turn out to follow from requiring crossing invariance \cite{SCHFlat}}. In the expansion  \eqref{mtildebetaz1} above, these show up as poles at $\rho = 1$. We therefore expand around $\rho = 1$ and separate out these poles to get the following useful form.
\begin{equation} 
\label{betarescsdrrho}
\begin{split}
&\text{Res}_{\b = -2n} \widetilde{M}(\b,z) \equiv \widetilde{\mathcal{W}}(n,\r) = \sum_{\t,\ell, k}^{\infty} \tau_k^{-n-1} c^{(k)}_{\t,\ell}\bigg[ \sum_{q=1}^{n-3} \frac{N_{q}(n, \t,\ell, k)}{(\rho-1)^{q}} + \mathcal{F}_{B}(n, \t,\ell, k, \rho)\bigg]
\end{split}
\end{equation}
$\mathcal{F}_{B}(n,\t,\ell, k, \rho)$ is a polynomial of degree ${\lfloor \frac{n}{2} \rfloor}$ in $\rho$ . We will refer to it as the \textit{Celestial Witten block}. Demanding that the unphysical poles at $\rho=1$ vanish gives the conditions
\begin{equation}
\label{locconstr}
\begin{split}
 \sum_{\t,\ell=2, k=0}^{\infty} \tau_k^{-n-1} c^{(k)}_{\t,\ell} {N_{b}(n, \t,\ell, k)} =0 \quad \text{for all } n \ge 4 \text{ and } b=1,2,\dots, n-3 \,.
\end{split}
\end{equation}
These conditions have been termed \textit{null/locality constraints} in the literature \cite{RGASAZ, SCHFlat}. We will refer to $N_{b}(n, \t,\ell, k)$ as \textit{Null blocks}. A compact formula for the Null blocks is as follows.
\begin{equation}
\label{cknJdef}
\begin{split}
N_{r}(n, \t,\ell, k) & = -  \frac{ 2^{\ell-3}}{(n-3-r)!}\  \lim_{\rho \to 1} \frac{d^{n-3-r}}{d\rho^{n-3-r}}\bigg[ (1+\rho)^{n-\ell}\mathcal{Q}_{\D,\ell}^{(k)}(-2n,\rho)\bigg] 
\end{split}
\end{equation}
In section 5, we show that they satisfy positivity properties that imply a low-spin dominance for AdS EFTs. \\
After imposing the null constraints, we equate \eqref{eq:LowEnergyRes} and \eqref{betarescsdrrho} to get a partial wave expansion for the Wilson coefficients as follows
\be
\label{PWE}
\sum_{\substack{p,q \\ 2p+3q=n}}\frac{(-1)^q}{2^{p+q}} \mathcal{W}_{p,q} 
(1+\rho)^{q} (1-\rho)^{p} =    \sum_{\t,\ell, k}\tau_k^{-n-1} c^{(k)}_{\t,\ell}  \mathcal{F}_{B}(n,  \t,\ell, k, \rho)
\ee
\section{Bounds on AdS EFTs and the HPPS conjecture}
In this section, we will examine the positivity properties of the Celestial Witten blocks. As mentioned earlier, they are polynomials of degree ${\lfloor \frac{n}{2} \rfloor}$ in $\rho$. 
\subsection{Positivity properties of the Celestial Witten Blocks}
We perform numerical checks and find that the Celestial Witten blocks satisfy the following positivity properties.
\begin{enumerate}
\item \textbf{Sign Definiteness}: \textit{For any $m$, the coefficients of $\rho^m$ in $\mathcal{F}_{B}(n,  \t,\ell, k, \rho)$ have the same sign for all $\ell\geq \ell_{SD}(n)$} and above some critical twist $\tau_{SD}(n,\ell)$. \\
For example, we find $\ell_{SD}(n=4,5)=4$. $\tau_{SD}(n,\ell)$ in general depends on the spacetime dimension $d$ of the CFT and the scaling dimension $\D_{\phi}$. For $d = 4,\D_{\phi} = 4$, we find for $n=4$, $\tau_{SD}(4,4) \approx 8.6$, $\tau_{SD}(4,6)  \approx 7.5$ and $\tau_{SD}(4,8) \approx 7.1$. For $n=5$, $\tau_{SD}(5,4) \approx 11.3$, $\tau_{SD}(5,6) \approx  8.0$ and $\tau_{SD}(5,8)\approx 7.4$.
\item \textbf{Typically Realness}: A function $f(z)$ is said to be typically real within the unit disc $|z|<1$ if $ Im \left(f(z)\right) Im ( z) > 0, \forall z$ (except when $Im (z) = 0$).\\
We find that, \textit{ $\mathcal{F}_{B}(n,  \t,\ell, k, \rho)$ is a typically real polynomial of $\rho$ within the unit disc $|\rho|<1$ for all $\ell \geq \ell_{TR}(n)$} and above some critical twist $\tau >\tau_{TR}(n,\ell)$.\\
For example, $\ell_{TR}(4)=4$ and $\ell_{TR}(5)=6$. For $d = 4,\D_{\phi} = 4$, we find for $n=4$, $\tau_{TR}(4,4) \approx 8.2$, $\tau_{TR}(4,6) \approx 7.2$ and $\tau_{TR}(4,8) \approx 6.9$. For $n=5$, $\tau_{TR}(5,6) \approx 9.3$ and $\tau_{TR}(5,8) \approx 8.1$.
\end{enumerate}
In general, we refer to the larger of $\ell_{SD}$ and $\ell_{TR}$ as $\ell_c$ and the larger of $\tau_{SD}(n,\ell)$ and $\tau_{TR}(n,\ell)$ as $\tau^*(n,\ell)$. As we will show in the next section, we require the positivity properties above to derive bounds. This naturally leads us to consider CFTs where there are no operators of spin $\ell \ge \ell_c (n)$ below $\t^*(n,\ell)$. For such CFTs, the RHS of \eqref{PWE} can be separated into two parts as
\be
\begin{split}
\label{RHSPWE}
\sum_{\ell=0}^{\ell_c-2} \sum_{\t,k}^{\infty} \tau_k^{-n-1} c^{(k)}_{\t,\ell}  \mathcal{F}_{B}(n,  \t,\ell, k, \rho)+\underbrace{ \sum_{\ell=\ell_c}^{\infty} \sum_{\t > \t^*(n,\ell),k}^{\infty} \tau_k^{-n-1} c^{(k)}_{\t,\ell}  \mathcal{F}_{B}(n,  \t,\ell, k, \rho)}_{T^{(heavy)}_n(\rho)} \,.
\end{split}
\ee
$T^{(heavy)}_n(\rho)$ is designated so because it refers to a sum of only heavy, higher spin operators. It follows from unitarity (positivity of the OPE coefficients squared) that $T^{(heavy)}_n(\rho)$ is a positive sum of Celestial Witten Blocks. And since positive sums preserve both sign definiteness and typically realness (\cite{QFTEFTGFT}), $T^{(heavy)}_n(\rho)$ inherits both the properties of Celestial Witten Blocks mentioned above. \\ \\
More precisely, lets expand $T^{(heavy)}_n(\rho) = \sum_{m=0}^{\lfloor\frac{n}{2} \rfloor} a_m(n)\rho^m$. Then we get
\begin{enumerate}
\item \textbf{Sign definiteness:}
\textit{For any $m$, the coefficient of $\rho^m$ in $T^{(heavy)}_n(\rho)$ has the same sign as the coefficients of $\rho^m$ in $\mathcal{F}_{B}(n, \t,\ell, k, \rho)$ for any $\ell \ge \ell_c$}, i.e.
\be
\label{signbound}
 a_m(n) \text{ always have fixed signs}
\ee
\item \textbf{Typically realness:}
\textit{$T^{(heavy)}_n(\rho)$ is a typically real polynomial of $\rho$ within the unit disc $|\rho|<1$}.\\
Taylor expansion coefficients of typically-real polynomials satisy two-sided bounds called Suffridge bounds \cite{suffridge}. For degree 1 polynomials ,i.e. $n=2,3$, the Suffridge bounds do not exist. Here, we quote them for $n=4,5$, i.e. degree 2 polynomials. For general bounds, see \cite{SGPRAS}
\be 
\label{TRbound}
\Big|\frac{a_2(n)}{a_1(n)}\Big| \le \frac{1}{2}, \quad \text{for } n = 4,5
\ee
\end{enumerate}

\noindent \textbf{Why can we not have two-sided bounds for $\mathbf{n=2, 3}$, i.e. on $\mathbf{\mW_{1,0}}$ and $\mathbf{\mW_{0,1}}$?}\\
As mentioned in the Introduction, the results of \cite{Swamp, chuotAdS} suggest that we can only derive two-sided bounds on the ratio $\mW_{0,1}/\mW_{1,0}$, not on each Wilson coefficient. In our formalism, this follows from the fact that for $n =2$ and $n =3$, the Celestial Witten blocks are degree 1 polynomials given as $\mathcal{F}_{B}(2, \t,\ell, k, \rho) \sim (1 - \rho)$ and $\mathcal{F}_{B}(3, \t,\ell, k, \rho) \sim (1 + \rho)$. For degree 1 polynomials, the Suffridge bounds do not exist. Further, it is easy to see by plugging in $\mathcal{F}_{B}(2, \t,\ell, k, \rho)$ and $\mathcal{F}_{B}(3, \t,\ell, k, \rho)$ in \eqref{PWE} that sign definiteness also can only lead to one-sided bounds. In an upcoming work \cite{FBAZ_GFT}, we study the positivity properties of the full Mellin amplitude to show that two-sided bounds can be derived on the ratio $\mathcal{W}_{0,1}/\mathcal{W}_{1,0}$.

\subsection{Bounds on AdS Wilson Cofficients}
In this section, we will use the positivity properties discussed above to derive bounds on the AdS Wilson coefficients. These will be given in terms of discrete moments of OPE coefficients defined as follows
\be
\label{OPEMoments}
\widehat{C} _{\ell,m}(n) = \sum_{\t,k}^{\infty} \tau_k^{-n-1} c^{(k)}_{\t,\ell}b_m(n,\ell,k)
\ee
where $b_m(n,\ell,k)$ are the coefficients of $\rho^m$ in the expansion of the Celestial Witten Blocks. The moments are always finite. The sum over $k$ converges because the normalization factor in $c^{(k)}_{\t,\ell}$ falls off faster than $k^{-1} e^{-\frac{1}{k}}$ at large $k$ (see \eqref{DExp}). The sum over $\t$ is also convergent because the OPE coefficient for holographic theories is known to fall off exponentially at large $\t$ \cite{RychkovOPE}.\\
In terms of the moments, equation \eqref{RHSPWE} reads
\begin{eqnarray}
\label{TnrhoDef}
T_n^{(heavy)}(\rho) =\left( \sum_{\substack{p,q \\ 2p+3q=n}}\frac{(-1)^q}{2^{p+q}} \mathcal{W}_{p,q} 
(1+\rho)^{q} (1-\rho)^{p} \right)-\sum_{\ell=0}^{\ell_c-2} \widehat{C}_{\ell,m}(n)\rho^m
\end{eqnarray}
\\
We can now simply expand both sides in powers of $\r$ and use the positivity properties of $T_n^{(heavy)}(\rho)$ given in \eqref{signbound} and \eqref{TRbound} to derive bounds on $\mathcal{W}_{p,q}$. We label the bounds following from sign definiteness as SD and those from typically realness as TR. Let us see how this works case-by-case starting from $n=2$. 

\subsubsection*{$\mathbf{n = 2 \text{ and } n = 3 }$ }
For $n=2$, we simply get a lower bound
\be 
\mathbf{SD:} \quad \mathcal{W}_{1,0} \ge  0
\ee
For $n = 3$, we get
This leads to a lower bound
\begin{eqnarray}
\mathbf{SD:}  \quad \mathcal{W}_{0,1} \ge   -2\widehat{C}_{0,0}(3)
\end{eqnarray}
\subsubsection*{$\mathbf{n = 4}$}
For $n=4$, $\ell_c = 4$ and we find that $a_0(4) \le 0$, $a_1(4) \ge 0$ and $a_2(4) \le 0$. This leads to the following bounds
\begin{eqnarray}
\label{n4CFTbound1}
\mathbf{SD:}  \quad  \frac{1}{4} \mathcal{W}_{2,0} \le  \widehat{C}_{0,0}(4)+ \widehat{C}_{2,0}(4), \quad \frac{1}{2} \mathcal{W}_{2,0} \ge  \widehat{C}_{0,1}(4)+ \widehat{C}_{2,1}(4) , \quad \frac{1}{4} \mathcal{W}_{2,0} \ge \widehat{C}_{0,2}(4)+ \widehat{C}_{2,2}(4)
\end{eqnarray}
From typically realness, we get
\begin{equation}
\label{n4CFTbound2}
 \begin{split}
 \mathbf{TR:}\quad
-\frac{1}{2} &\le \frac{\frac{1}{4} \mathcal{W}_{2,0} - \widehat{C}_{0,2}(4)- \widehat{C}_{2,2}(4)}{\frac{1}{2} \mathcal{W}_{2,0} - \widehat{C}_{0,1}(4)- \widehat{C}_{2,1}(4)} \le \frac{1}{2}\,,  \\
\implies \mathcal{W}_{2,0} &> 2 \widehat{C}_{0,2}(4)+ 2\widehat{C}_{2,2}(4)+\widehat{C}_{0,1}(4)+ \widehat{C}_{2,1}(4)
\end{split}
\end{equation}
\subsubsection*{$\mathbf{n = 5}$}
For $n=5$, $\ell_c = 6$ and we find that $a_0(5) \le 0$  and $a_2(5) \ge 0$. This leads to the following bounds
\begin{eqnarray}
\label{n5CFTbound1}
 \mathbf{SD:} \quad \frac{1}{4} \mathcal{W}_{1,1} \ge \widehat{C}_{0,0}(5)+ \widehat{C}_{2,0}(5)+\widehat{C}_{4,0}(5),  \quad \frac{1}{4} \mathcal{W}_{1,1} \ge \widehat{C}_{0,2}(5)+ \widehat{C}_{2,2}(5)+\widehat{C}_{4,2}(5)
\end{eqnarray}
From typically realness, we get
\begin{equation}
\label{n5CFTbound2}
\begin{split}
 \mathbf{TR:} &\quad  
-\frac{1}{2} \le \frac{\frac{1}{4} \mathcal{W}_{1,1} - \widehat{C}_{0,2}(5)- \widehat{C}_{2,2}(5)-\widehat{C}_{4,0}(5)}{ - \widehat{C}_{0,0}(5)-  \widehat{C}_{2,0}(5)-\widehat{C}_{4,0}(5)}  \le \frac{1}{2}  \\
\implies \quad \mathcal{W}_{1,1} &\ge 4 \widehat{C}_{0,2}(5)+  4\widehat{C}_{2,2}(5)+  4\widehat{C}_{4,2}(5)-2\widehat{C}_{0,0}(5)- 2\widehat{C}_{2,0}(5) - 2\widehat{C}_{4,0}(5) \quad \text{and}\\
 \mathcal{W}_{1,1} &\le 4 \widehat{C}_{0,2}(5) + 4\widehat{C}_{2,2}(5)+4 \widehat{C}_{4,2}(5) +2\widehat{C}_{0,0}(5)+ 2\widehat{C}_{2,0}(5)+ 2\widehat{C}_{4,0}(5)
\end{split}
\end{equation}
Similarly, we can go to higher $n$. In each case, sign definiteness and typically realness of Celestial Witten blocks prove sufficient to derive two-sided bounds.

\subsubsection{Connection to the HPPS conjecture}
We found above that to derive the two-sided bounds on Wilson coefficients, it is sufficient that all spin 4 and higher operators appear only above some critical twist $\text{max}(\t^*(n,\ell))$. This leads us to the statement \textit{``Any CFT in which all spin four and higher operators have twists greater than some critical} $\text{max}(\t^*(n,\ell))$\textit{ has all Wilson Coefficients beginning from $\mW_{2,0}$ onwards bounded on both sides".} We therefore prove in a precise wasy, a form of the HPPS conjecture which states that similar conditions are sufficient for a (large N) CFT to have a local AdS dual.

\section{Bounds in the flat space limit}
In this section, we derive bounds on AdS Wilson coefficients in a $1/R$ expansion, i.e. around the flat-space limit \cite{Penedones:2010ue}.  The details of taking the flat-space limit change depending on the space-time dimension and the dimension $\D_{\phi}$ of external scalars. Our procedure is independent of these details. Our goal is to check that the bounds we derive are respected by the AdS Virasoro-Shapiro amplitude \cite{Alday1, Alday2}. So, in this section, we will focus on Mellin amplitudes of scalars with $\D_{\phi}=4$ in $d=4$. We consider Mellin amplitudes with the following large $R$ expansion.  \\
\be\label{Mexp}
M(s_1,s_2) = 8 \pi G M_{sugra} + \frac{1}{R^{6}}\sum_{p,q} \frac{\Gamma(2p+3q+6)}{(R^{2})^{2p+3q}} x^p y^q \left(\mW^{(0)}_{p,q}+ \frac{\mW^{(1)}_{p,q}}{R^2}+...\right)
\ee
The above normalization of AdS Wilson coefficients is chosen so that in the flat-space limit, we recover the Wilson coefficients of the corresponding flat-space scattering amplitude.
\be\label{MexpFlat}
\mathcal{M}_{flat}(S_1,S_2) = \frac{8 \pi G}{S_1 S_2 S_3}+ 2\sum_{p,q} \mW^{(0)}_{p,q} X^p Y^q 
\ee
$S_1, S_2$ are the Mandelstam variables and $X, Y$ are the flat-space analogues of $x,y$. The normalization follows from the the \textit{flat-space transform} that relates the CFT Mellin amplitude and the flat-space scattering amplitude. Explicitly (see \cite{Alday2}),
\be\label{FSTrans}
FS(M(s_1,s_2)) \sim \lim_{R \rightarrow \infty} R^6 \int_{-i \infty}^{+i \infty} \frac{d\a}{2\pi i} e^{\a}\a^{-6} M (\frac{R^2}{\alpha} s_1, \frac{R^2}{\alpha} s_2) = \mathcal{M}_{flat}(S_1,S_2)
\ee
The transform effectively removes the $\Gamma(2p+3q+6)$ factor in the expansion above without which the sum would not converge. \\
An important example that falls in the category of amplitudes in \eqref{Mexp} is the planar correlator of four stress-tensor multiplets in $\mathcal{N}=4$ SYM. This amplitude is the AdS analogue of the Virasoro-Shapiro amplitude and reduces to it in the flat space limit.
\be
\label{VSAmp}
\mathcal{M}_{flat}^{VS}(S_1,S_2)= \frac{8 \pi G}{S_1 S_2 S_3}\frac{\Gamma(1-S_1)\Gamma(1-S_2)\Gamma(1+S_1+S_2)}{\Gamma(1+S_1)\Gamma(1+S_2)\Gamma(1-S_1-S_2)}
\ee
The coefficients $\mW^{(0)}_{p,q}$ for this amplitude can be easily found by  expanding $\mathcal{M}_{flat}^{VS}(S_1,S_2)$ as in \eqref{MexpFlat}. In \cite{Alday2}, the sub-leading curvature corrections $\mW^{(1)}_{p,q}$ and their partial wave expansion were fully determined as well.
\\
We'll write the Celestial Witten block expansion \eqref{PWE} in a $1/R$ expansion and use it to put bounds on $\mW^{(0)}_{p,q}$ and in some cases, also on $\mW^{(1)}_{p,q}$. Our bounds on $\mW^{(0)}_{p,q}$ exactly match those of \cite{SGPRAS} in flat-space, while the bounds on $\mW^{(1)}_{p,q}$ are new. We compare our results with the $1/R$ expansion of the AdS Virasoro-Shapiro amplitude and find that it satisfies our bounds upto order $1/R^2$. 

\subsection{1/R expansion of the Conformal Partial Wave Expansion}
We define $\mW_{p,q} (R) \equiv \mW^{(0)}_{p,q}+ \mW^{(1)}_{p,q}/R^2+\dots$ as the AdS Wilson coefficient. The goal is to write the RHS of \eqref{PWE} in a $1/R$ expansion. Since we are going to eventually compare with a supersymmetric amplitude, we will work with a partial wave expansion with a twist shifted by 4. This is to take into account the factorization of supersymmetric amplitudes due to the superconformal Ward Identity. We then have 
\be
\label{PWESUSY}
\begin{split}
\sum_{\substack{p,q \\ 2p+3q=n}}\frac{(-1)^q}{2^{p+q}} \mathcal{W}_{p,q} 
(1+\rho)^{q} (1-\rho)^{p} =    \sum_{\t,\ell, k}(\tau_k+4)^{-n-1} c^{(k)}_{\t+4,\ell} \mathcal{F}_{B}(n,  \t + 4,\ell, k, \rho)
\end{split}
\ee
The flat space limit from the CFT side is controlled by the fixed $\ell$, large $\t$ limit. Following \cite{Alday2}, we therefore parametrize $\t$ and the OPE coefficients as
\begin{equation}
\label{LargeRPara}
\begin{split}
\t(r,\ell,R) = m_0(r) R &+ m_1(r,\ell)+ \frac{m_2(r,\ell)}{R}+ \dots \,,\quad
c^{(k)}_{\t+4,\ell}(r,R) = \mathcal{D}(\t,\ell,k)f(r,\ell,R),\\
&f(r,\ell,R) =  f_0(r,\ell)+\frac{f_1(r,\ell)}{R}+\frac{f_2(r,\ell)}{R^2}+ \dots\,
\end{split}
\end{equation}
where $r$ denotes the collection of all other quantum numbers characterising the CFT operators, $f_0(r,\ell)$ are related to the flat space partial wave coefficients (see \eqref{aTof}) and $\mathcal{D}(\t,\ell,k)$ is a normalization factor (see \eqref{DDef}). \\
We are left with the job of expanding the Witten blocks and the normalization factor $\mathcal{D}(\t,\ell,k)$. This can done easily by first expanding in large $\t$, plugging the expansion of $\t(r,\ell,R)$ and expanding again in large $R$. For the Witten block, this computation is involved. We provide an expansion of the Mack polynomial in the large $s$ and large $\nu = \tau + \ell -h$ limit in the Appendix \eqref{macklarges}  that makes it simpler. \\
\\
The final step is to perform the $k$ sum. It is carried out by first noticing that for large $\tau$, the dominant contribution must come from terms of order $k \sim \tau^2$. We can thus define $k = x \t^2$ and turn the $k$ sum into an integral,  $\sum_k \rightarrow \t^2 \int_{0}^{\infty}dx$ following \cite{Alday1, Alday2}. The integral is easy to perform term by term in $1/R$.  
\subsubsection*{Constraints from AdS}
The expansion of the amplitude/Wilson coefficients in \eqref{Mexp} has no correction proportional to odd powers of $1/R$. This is not automatic from the partial wave expansion \eqref{PWESUSY}. Imposing it for the first odd power fixes $m_1(r,\ell)$ and  $f_1(r,\ell)$ in terms of $m_0(r)$ and $f_0(r, \ell)$. Since there are just two unknowns to be fixed, we can demand that first odd power vanishes for $n=4,5$ and it will hold true for higher $n$ as well. This yields 
\be
m_1(r,\ell) = -\ell - 2, \quad f_1(r,\ell) =\frac{23 + 12\ell}{2 m_0 (r)} f_0(r,l)
\ee
The same result is obtained in \cite{Alday1}.
\subsection{Bounds in the 1/R expansion}
\subsubsection{Bounds in terms of partial wave moments}
In the flat space limit, the Mack polynomial reduces to the Gegenbauer polynomial (see \eqref{macklarges}). This means that the Celestial Witten blocks reduce to crossing symmetric and local blocks given in terms of the Gegenbauer polynomials. 
These blocks were termed Celestial Feynman blocks in \cite{SGPRAS} and were shown to satisy similar positivity and typically realness properites as we find for the Celestial Witten blocks. For exact results on the positivity properties of Celestial Feynman blocks, we refer the reader to \cite{SGPRAS}. At the sub-leading order, sign-definiteness is not guaranteed but we find that typically realness continues to hold. \\
Instead of the OPE moments $\widetilde{C}_{l,m}(n)$, we get moments of $f_0(r,l)$ and the correction $f_2(r,l)$. These are defined as
\begin{eqnarray}
\begin{split}
\phi^{(0)}(\ell,n) &= \sum_r \frac{f_0(r,\ell)}{(m_0(r)/2)^{2n+6}},\quad \phi^{(1)}_1(\ell,n) = \sum_r \frac{f_0(r,\ell)\t_2(r,\ell)}{(m_0(r)/2)^{2n+6+1}},\\
\phi^{(1)}_2(\ell,n) &= \sum_r \frac{f_0(r,\ell)}{(m_0(r)/2)^{2n+8}}, \quad \phi^{(1)}_3(\ell,n) = \sum_r \frac{f_2(r,\ell)}{(m_0(r)/2)^{2n+6}}
\end{split}
\end{eqnarray}
$\phi^{(0)}(\ell,n)$ only appears at the leading order in $1/R$ and is positive due to flat space partial wave unitarity. The rest occur at the sub-leading order in $1/R$.
\subsubsection*{$\mathbf{n=2, 3}$}
For $n = 2$, at the leading order in $1/R^2$, we find
\be
\mathbf{SD:}  \quad   \phi^{(0)}(0,2) \le \mathcal{W}^{(0)}_{1,0} 
\ee
For $n = 3$, we find 
\be
\mathbf{SD:}  \quad  -\frac{3}{2}\phi^{(0)}(0,3) \le \mathcal{W}^{(0)}_{0,1}  
\ee
For AdS Virasoro-Shapiro, this leads to the bounds  $\mathcal{W}^{(0)}_{1,0} \ge 1$ and $\mathcal{W}^{(0)}_{0,1} \ge -1.506$ while the actual values are $\mathcal{W}^{(0)}_{1,0} = 1.037$ and $\mathcal{W}^{(0)}_{0,1} = -1.445$.
\subsubsection*{$\mathbf{n=4}$}
At the leading order in $1/R^2$, we find 
\begin{eqnarray}
\begin{split}
\mathbf{SD:}  \quad  
 & \phi^{(0)}(0,4)+ \phi^{(0)}(2,4) \le \mathcal{W}^{(0)}_{2,0}  \le \phi^{(0)}(0,4) + 17\phi^{(0)}(2,4)\\
\mathbf{TR:}  \quad   &\mathcal{W}^{(0)}_{2,0} \ge \phi^{(0)}(0,4)- \frac{1}{3}\phi^{(0)}(2,4)
\end{split}
\end{eqnarray}
For AdS Virasoro-Shapiro, this gives the bound $ 1.0078 \le \mathcal{W}^{(0)}_{2,0} \le 1.1015$ while the actual value is $\mathcal{W}^{(0)}_{2,0} = 1.0083$. \\
\textit{\textbf{Sub-leading order}}: We find
\begin{eqnarray}
\begin{split}
\mathbf{TR:}  \quad   \mathcal{W}^{(1)}_{2,0} &\le  \phi^{(1)}_3(0,4)- \frac{1}{3} \phi^{(1)}_3(2,4)
-7(\phi^{(1)}_1(0,4)-\frac{1}{3}\phi^{(1)}_1(2,4))\\
&-\Big(\frac{3967}{96}\phi^{(1)}_2(0,4)-\frac{11791}{288}\phi^{(1)}_2(2,4)\Big)
\end{split}
\end{eqnarray}
For AdS Virasoro-Shapiro, this gives the bound  $\mathcal{W}^{(1)}_{2,0} \le -40.266$ while the actual value is $\mathcal{W}^{(1)}_{2,0} = -40.302$.

\subsubsection*{$\mathbf{n=5}$}
At the leading order in $1/R^2$, we find 
\begin{eqnarray}
\begin{split}
\mathbf{SD:}  \quad  &-\frac{5 }{2}\phi^{(0)}(0,5) + \frac{17}{6} \phi^{(0)}(2,5) \le \mathcal{W}^{(0)}_{1,1} \\
\mathbf{TR:}  \quad  &-\frac{5 }{2}\phi^{(0)}(0,5) + \frac{1}{6} \phi^{(0)}(2,5) \le \mathcal{W}^{(0)}_{1,1} \le -\frac{5 }{2}\phi^{(0)}(0,5) + \frac{11}{2} \phi^{(0)}(2,5)
\end{split}
\end{eqnarray}
For AdS Virasoro-Shapiro, this gives the bound $  -2.4941 \le \mathcal{W}^{(0)}_{1,1} \le -2.4863$ while the actual value is $\mathcal{W}^{(0)}_{1,1} = -2.4929$. \\
\textit{\textbf{Sub-leading order}}: We find
\begin{eqnarray}
\begin{split}
\mathbf{TR:}  \quad  
\mathcal{W}^{(1)}_{1,1} &\le  -\frac{5}{2}\phi^{(1)}_3(0,5)+\frac{1}{6} \phi^{(1)}_3(2,5)
+(20\phi^{(1)}_1(0,5)-44 \phi^{(1)}_1(2,5))\\
&+\Big(\frac{10185}{64}\phi^{(1)}_2(0,5)-\frac{210607}{576}\phi^{(1)}_2(2,5)\Big)\\
\mathbf{TR:}  \quad   \mathcal{W}^{(1)}_{1,1} &\ge  -\frac{5}{2}\phi^{(1)}_3(0,5)+\frac{11}{2} \phi^{(1)}_3(2,5)
+(20\phi^{(1)}_1(0,5)-\frac{4}{3} \phi^{(1)}_1(2,5))\\
&+\Big(\frac{10185}{64}\phi^{(1)}_2(0,5)-\frac{50639}{576}\phi^{(1)}_2(2,5)\Big)\\
\end{split}
\end{eqnarray}
For AdS Virasoro-Shapiro, this gives the bound  $ 161.3481 \le \mathcal{W}^{(1)}_{1,1} \le 161.5155$ while the actual value is $\mathcal{W}^{(1)}_{1,1} = 161.4318$.

\subsubsection{Bounds in terms of the mass gap}
In terms of the flat space partial wave coefficients (see Appendix \eqref{AppC} for our conventions), we can write $\phi^{(0)}(n,\ell) = 2^8 (\ell+1)^2 \int_{4M^2}^{\infty} \frac{ds}{s^{n+1+1/2}} a_{\ell}(s)$. Using partial wave unitarity ($0 \le  a_{\ell}(s) \le 1$), this leads to the bounds $0 \le \phi^{(0)}(n,\ell) \le \frac{2^9 (\ell+1)^2}{(2n+1)(4M)^{2n+1}}$. This allows us to write the flat space bounds given before in terms of the mass gap $M$. We get
\be
\begin{split}
\frac{3.2}{M^5} \le \mathcal{W}^{(0)}_{1,0}, \qquad -\frac{0.857}{M^5} \le \mathcal{W}^{(0)}_{1,0}, \qquad 0 \le \mathcal{W}^{(0)}_{2,0} \le \frac{17.111}{M^9}, \qquad
-\frac{0.057}{M^{11}} \le \mathcal{W}^{(0)}_{1,1} \le \frac{1.125}{M^{11}}
\end{split}
\ee
Note that these are bounds on the Wilson coefficients themselves, not on the ratios of two Wilson coefficents or ratio with respect to $G$. The scaling with the mass gap therefore depends on the spacetime dimension $D$ of flat space  as $\mW_{p,q} \sim M^{4-2n-D},~ n = 2p+3q$.

\section{Low-Spin Dominance for AdS Amplitudes}
\begin{figure}[hbt!]
     \centering
     \begin{subfigure}[b]{0.45\textwidth}
         \centering
         \includegraphics[width=\textwidth]{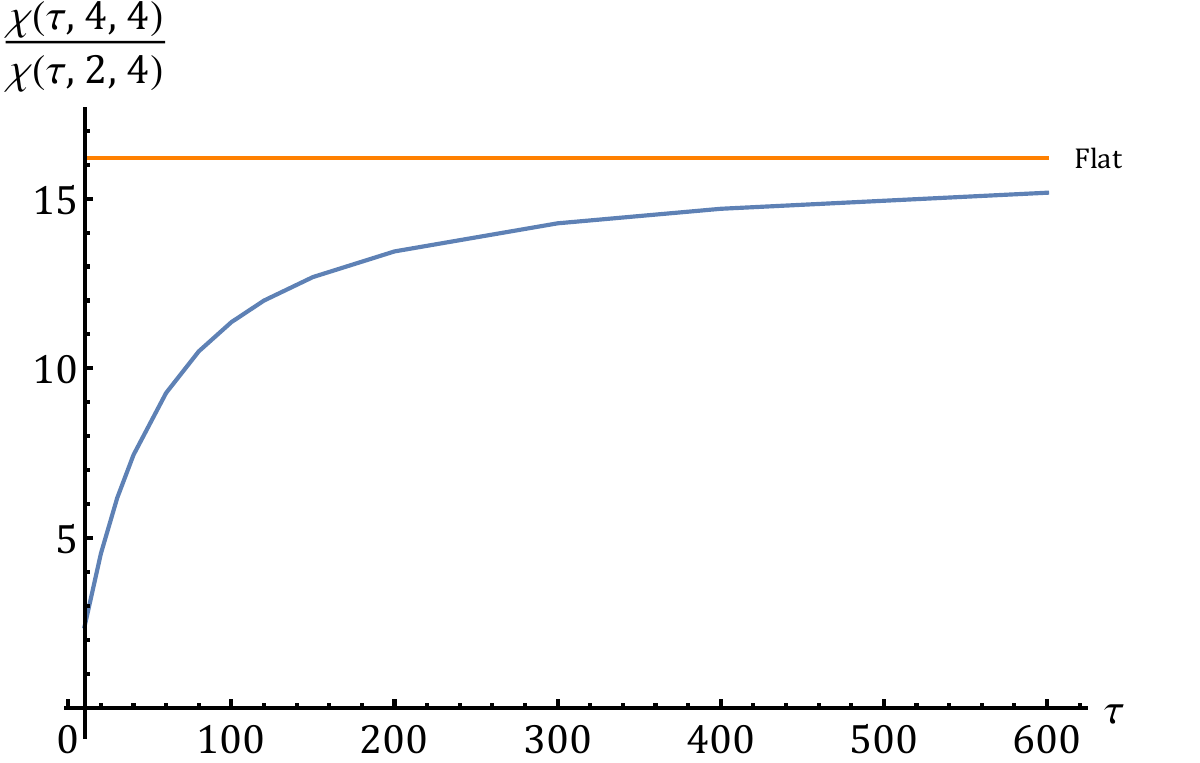}
      \end{subfigure}
     \begin{subfigure}[b]{0.45\textwidth}
         \centering
         \includegraphics[width=\textwidth]{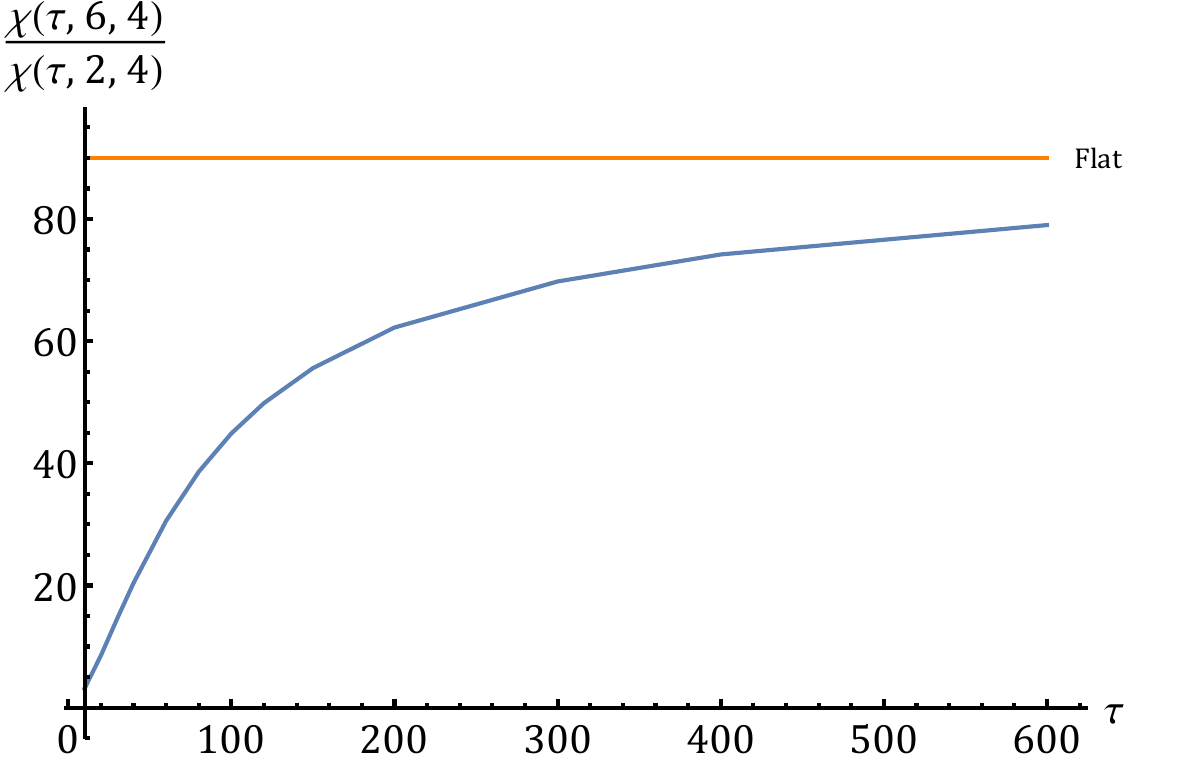}
      \end{subfigure}
\caption{The plots suggest that low-spin dominance weakens as we move away from the flat space limit ($\t \rightarrow \infty$)}
\label{fig:LSDPLOT}
\end{figure}

In this section, we explore the consequences of the null constraints given in \eqref{locconstr}. Considering the first null constraint, we find that\\\textit{``For $n=4$, the spin two Null blocks are always positive while the higher spin Null blocks are always negative beyond some critical twist"}.\\
The critical twist here is not the same as the critical twist we found for positvity of $\ell \ge 4$ Celestial Witten blocks before and in general, we denote the larger of the two by $\t^*(4,\ell)$. Considering CFTs where all $\ell \ge 4$ operators are above $\t^*(4,\ell)$, we get the following relation
\begin{equation}
\label{LSDCFT}
\begin{split}
 \sum_{\t,\ell=2, k}^{\infty} \tau_k^{-5} c^{(k)}_{\t,\ell} {N_{1}(4, \t,\ell, k)} = \sum_{\ell=4}^{\infty}\sum_{\t > \t^*(4,\ell),k}^{\infty} \tau_k^{-5} c^{(k)}_{\t,\ell} {N_{1}(4, \t,\ell, k)}
\end{split}
\end{equation}
This suggests that the spin two contribution dominates any other higher spin contribution in the CFT. To quantify this dominance, we again focus us the case when $d=4,  \D_{\phi} = 4$ although the conclusions we derive hold generally. We parametrize the twists and OPE coefficients as in \eqref{LargeRPara} and look at the leading order behaviour in $1/R$. The shift $\t \rightarrow \t + 4$ is not significant here as we are interested in the large $\t$ behaviour. In the flat space limit, we recover
\be
\frac{16}{3}\phi^{(0)}(2,4) = \frac{432}{5}\phi^{(0)}(4,4)+ 480\phi^{(0)}(6,4)+ \dots
\ee
This implies a low-spin dominance for flat space partial wave moments as follows
\be
\label{LSDFlat}
\frac{\phi^{(0)}(4,4)}{\phi^{(0)}(2,4)} \ge 16.2, \quad \frac{\phi^{(0)}(6,4)}{\phi^{(0)}(2,4)} \ge 90
\ee
It is interesting to study how low-spin dominance changes as we move slightly away from flat space. To make this comparison, we plug $c^{(k)}_{\t,\ell} = \mathcal{D}(\t,\ell,k)f(\t,\ell)$ and define 
\be
\chi(\t,\ell,4) = \sum_{k}  \frac{\mathcal{D}(\t,\ell,k) N_{1}(4, \t,\ell, k)}{\t_k^5}
\ee
From \eqref{LSDCFT}, it follows that
\be
\sum_{\t} f(\t,2)\chi (\t,2,4) \ge \sum_{\t > \t^*(4,\ell)}  f(\t,\ell)\chi (\t,\ell,4) \quad \text{for all } \ell \ge 4
\ee
We are interested in the large $\t$ behaviour of the ratios $\chi(\t,4,4)/\chi(\t,2,4)$ and $\chi(\t,6,4)/\chi(\t,2,4)$. We plot these in figure \eqref{fig:LSDPLOT} and find that they increase with $\t$ and converge towards their maximum values in the flat space ($\t \rightarrow \infty$) limit (as given in \eqref{LSDFlat}). This suggests that \\
\textit{``Low-spin dominance of partial wave coefficients is the strongest in flat space and weakens as we move away from flat space towards higher AdS curvature"}

\section{Discussion}\label{discuss}
In this paper, we introduce a new approach to study CFT Mellin amplitudes motivated by the program of Celestial amplitudes in flat space. We perform a Celestial transform of the CFT Mellin amplitude. Using the crossing symmetric dispersion relation (CSDR), we expand the resulting amplitude in a basis of crossing symmetric partial waves which we refer to as the Celestial Witten blocks. The blocks possess remarkable positivity properties in the variable $\rho$ which is related to the Celestial variable $z$. These positivity properties along with the positivity of OPE coefficients squared (unitarity) are sufficient to derive two-sided bounds on Wilson coefficients.\\
 \\
The positivity properties we derive are sufficient to derive two-sided bounds on all Wilson coefficients except $\mathcal{W}_{1,0}$ and $\mathcal{W}_{0,1}$ for which we only find a lower bound. In an upcoming work \cite{FBAZ_GFT}, we will show that the ratio $\mathcal{W}_{0,1}/\mathcal{W}_{1,0} $ can also be bounded. This follows from positivity properties of the full Mellin amplitude which imply two-sided bounds on ratios of all Wilson coefficients in a CFT. \\ 
\\
We then turn to deriving bounds in a $1/R$ expansion around the flat space limit. The goal is to compare our results with the AdS Virasoro amplitude which is given by the planar Mellin correlator of four stress tensor multiplets in $\mathcal{N}=4$ SYM. We find that it satisfies our bounds upto sub-leading order in $1/R^2$. A complication that arises in taking the flat space limit is the appearance of the graviton pole $\frac{x^2}{y}$ in the forward scattering limit. Due to this pole, the partial wave expansion we write does not hold in the flat space limit for Wilson coefficients $\mW_{p,q}$ with $2p + 3q < 4$, in particular, $\mathcal{W}_{1,0}$ and $\mathcal{W}_{0,1}$. We do not face this issue in the example we study because we consider a supersymmetric amplitude where the graviton pole appears as $\frac{1}{y}$. In an upcoming work, we show how to deal with the graviton pole $\frac{x^2}{y}$ in a simple manner using the CSDR \cite{FBAZ_Gravity}.\\ \\
In the future, it will be interesting to extend our methods to theories with spinors as well as theories with  global symmetries \cite{Apratim, Charge,Chowdhury:2021qfm}. Another area of application is the pion bootstrap at large $N$ \cite{FBAZ_Large_N_QCD, AZ, RastelliPion, Pion2, gpv}. It will also be interesting to examine the consequences of our methods for weakly-coupled CFTs\cite{FBRG,RGG}. QFTs in dS background as considered in \cite{cosmo} offer an intriuging area of applicaton as well. Another interesting direction is to study the connection between typically realness and the positive geometry of scattering amplitudes \cite{positivegeo}.

\section*{Acknowledgments}
We thank Aninda Sinha for initial collaboration and numerous helpful discussions.  We especially thank Tobias Hansen for clarifications on the flat space limit. We thank Sudip Ghosh and Prashant Raman for various discussions. We also thank Parthiv Halder and Balt Van Rees for their comments on the draft. We acknowledge partial support from  DST through the SERB core grant CRG/2021/000873.

\appendix

\section{Details of conventions} \label{AppA}
We define $c^{(k)}_{\t,\ell}=C_{\D,\ell}\mathcal{N}_{\D,\ell}\mathcal{R}_{\D,\ell}^{(k)}$, where $C_{\D,\ell}$ is the OPE coefficient squared as defined in \cite{RGASAZ} and the normalization factor is given as
\be\nonumber
\begin{split}
\mathcal{N}_{\Delta, \ell}=\frac{2^{\ell}(\Delta+\ell-1) \Gamma^{2}(\Delta+\ell-1) \Gamma(\Delta-h+1)}{\Gamma(\Delta-1) \Gamma^{4}\left(\frac{\Delta+\ell}{2}\right) \Gamma^{2}\left(\Delta_{\phi}-\frac{\Delta-\ell}{2}\right) \Gamma^{2}\left(\Delta_{\phi}-\frac{2 h-\Delta-\ell}{2}\right)}\,,\mathcal{R}_{\D,\ell}^{(k)}=\frac{\Gamma^{2}\left(\frac{\Delta+\ell}{2}+\Delta_{\phi}-h\right)\left(1+\frac{\Delta-\ell}{2}-\Delta_{\phi}\right)_{k}^{2}}{k ! \Gamma(\Delta-h+1+k)}\,.
\end{split}
\ee
The form for the Mack polynomial we use is \cite{RGASAZ}
\be
P^{(s)}_{\D-h,\ell}(s,t)=\sum_{m=0}^\ell\sum_{n=0}^{\ell-m}\mu^{(\D,\ell)}_{n,m}~~\left(\frac{\D-\ell}{2}-s\right)_{m}~~\left(-t\right)_{n}\,,
\ee
where $\m$ has a general closed form
\be\label{Ap:mu}
\begin{split}
&\mu^{(\D,\ell)}_{n,m}
=\frac{2^{-\ell} \ell! (-1)^{m+n} (h+\ell-1)_{-m} \left(\frac{\ell+\Delta }{2}-m\right)_m (\ell+\Delta -1)_{n-\ell} \left(\frac{\Delta -\ell}{2}+n\right)_{\ell-n} \left(\frac{\Delta -\ell}{2}+m+n\right)_{\ell-m-n} \,}{m! n! (\ell-m-n)!}\\
&{}_4F_3\left(-m,-h+\frac{\Delta-\ell }{2}+1,-h+\frac{\Delta -\ell}{2}+1,n+\Delta -1;\frac{\D+\ell}{2}-m,\frac{\D-\ell}{2}+n,-2 h-\ell+\Delta +2;1\right)\,.
\end{split}
\ee
In the main text, we use 
$$
P_{\D,\ell}(s_1,s_2)=P^{(s)}_{\D-h,\ell}(s_1+\frac{2\Dphi}{3},s_2+\frac{\Dphi}{3})\,.
$$
and we further define
\be
\mathcal{P}_{\D,\ell}^{(k)}\left(z\right)\equiv P_{\Delta, \ell}\left(\tau_{k}, s_{2}= \frac{\tau_k (z-1)}{2}\right)
\ee

\section{Flat Space Limit}  \label{AppB}
\subsection{Large R expansion of the Mack Polynomial}
The large $s$ and large $\nu = \tau + \ell -h $ expansion of the Mack polynomial is given as follows
\be
\begin{split}\label{macklarges}
& \mathcal{P}_{\t+\ell,\ell}^{(k)}(z)= \frac{8^{-\ell} \ell! s^\ell}{(h-1)_\ell}\Big[\left(C_\ell^{(h-1)}\left(1+\frac{2 t}{s}\right)-\frac{(h-1)}{\nu^2  }C_{\ell-2}^{(h)}\left(1+\frac{2 t}{s}\right)\right) \\
&+\frac{2 (h-1)^2}{s}\left( C_{\ell-2}^{(h)}\left(1+\frac{2 t}{s}\right)-\frac{1}{\nu^2}\left(h C_{\ell-4}^{(h+1)}\left(1+\frac{2 t}{s}\right)-C_{\ell-2}^{(h)}\left(1+\frac{2 t}{s}\right)\right)\right)\\
&+ \frac{h-1}{24 s^2}\Big(2\left(-3 h^2+2 h (5 \ell-7)+\ell^2-22\ell+31\right) C_{\ell-2}^{(h)}\left(1+\frac{2 t}{s}\right) \\
&+ 2h \left(24 h^2-40 h+1\right) C_{\ell-4}^{(h+1)}\left(1+\frac{2 t}{s}\right)\\
&+\frac{1}{\nu^2}\Big((4 h \left(24 h^2-2 h (4\ell+17)-2\ell^2 + 14\ell-9\right) C_{\ell-4}^{(h+1)}\left(1+\frac{2 t}{s}\right)\\
&-6 \left(4 h^2-8 h+3\right) C_{\ell-2}^{(h)}\left(1+\frac{2 t}{s}\right)+h \left(-48 h^3+32 h^2+83 h+3\right) C_{\ell-6}^{(h+2)}\left(1+\frac{2 t}{s}\right)\Big)\\
&+ 6\nu^2  C_{\ell-2}^{(h)}\left(1+\frac{2 t}{s}\right)\Big)\Big]+\dots
\end{split}
\ee
where we put $t = s_2-\dfrac{\Delta_{\phi}}{3}$ $s = s_1 + \dfrac{2\Delta_{\phi}}{3}$, $s_2 = \dfrac{z-1}{2}s_1$, $s_1 = \t_k$. Then, we simply need to plug $k = x \t^2$ and the expansion for $\t(r,\ell,R)$ in \eqref{LargeRPara} and expand around large $R$.

\subsection{Large R parametrization of the OPE coefficient}
The normalization factor relating the OPE coefficient squared to the flat space partial waves in  \eqref{LargeRPara} is given as 
\be
\label{DDef}
\mathcal{D}(\t,\ell,k) = \frac{\pi ^3 4^{-\ell-6}\tau ^{6}}{ 4^{\tau} (\ell+1) \sin ^2\left( \frac{\pi \tau }{2}\right)} \mathcal{R}_{\tau+4+\ell,\ell}^{(k)} \mathcal{N}_{\tau+4+\ell,\ell}
\ee
Again, we plug $k = x \t^2$ and expand around large $\t$ first. At leading order, this gives
\be 
\label{DExp}
\mathcal{D}(\t,\ell,x \t^2) = \tau ^{-2 \ell}\left(\frac{2^{3\ell-7} e^{-\frac{1}{4 x}}  x^{-l-6}}{(l+1) \t^6} + O\left(\frac{1}{\t^7}\right)\right)
\ee
We then plug in the expansion for $\t(r,\ell,R)$ and expand around large $R$.

\section{Conventions for flat space partial wave coefficients} \label{AppC}
The $f_0(r,\ell)$ are defined by (see Appendix C of \cite{Alday2})
\be
Im\left(\mathcal{M}(S_1,S_2) \right) = \sum_{r,\ell}
\frac{f_0(r,\ell)}{(\ell+1)S_1^2 } \pi \delta\left( S_1 - \left(\frac{m_0(r)}{2}\right)^2\right)\mathcal{C}^{(1)}_{\ell}\left( 1+  \frac{2S_2}{S_1}\right)
\ee
In our convention,
\be
Im\left(\mathcal{M}(S_1,S_2) \right) =  128 \pi \sum_{\ell}(2\ell +2) a_{\ell}(S_1)S_1^{-\frac{1}{2}} \mathcal{C}^{(1)}_{\ell}\left( 1+  \frac{2S_2}{S_1}\right)
\ee
where $a_{\ell}(S_1)$ is the imaginary part of the flat space partial wace coefficient. This convention ensures $0 \le a_{\ell} \le 1$. \\
Comparing both gives,
\be 
\label{aTof}
a_{\ell}(S_1) = \frac{1}{256}\sum_r \frac{f_0(r,\ell)}{(\ell+1)^2 S_1^2} S_1^{1/2} \delta\left(S_1 - \left(\frac{m_0(r)}{2}\right)^2\right)
\ee
This allows us to define a moment for partial wave coefficients $a_{\ell}(S_1)$ as
\be 
\widetilde{a}(\ell,n) = \int_{4M^2}^{\infty} \frac{dS_1}{S_1^{n+1+\frac{1}{2}}} a_{\ell}(S_1) = \frac{\phi^{(0)}(\ell,n)}{256 (\ell+1)^2}
\ee

\end{document}